\documentclass[final,3p,times,onecolumn]{elsarticle}
\usepackage{graphicx}
\usepackage{amsmath}
\allowdisplaybreaks 
\def\bea#1\eea{\begin{align}#1\end{align}} 

\newcommand{\bef}{\begin{figure}[htb]\centering}
\newcommand{\eef}{\end{figure}}

\begin{document}

\begin{frontmatter}
\title{Collins azimuthal asymmetries of hadron production inside jets}
\author{Zhong-Bo Kang\fnref{label1,label2,label3}}
\ead{zkang@physics.ucla.edu}
\author{Alexei Prokudin\fnref{label4,label5}}
\ead{prokudin@jlab.org}
\author{Felix Ringer\fnref{label6}}
\ead{fmringer@lbl.gov}
\author{Feng Yuan\fnref{label6}}
\ead{fyuan@lbl.gov}
\address[label1]{Department of Physics and Astronomy, University of California, Los Angeles, California 90095, USA}
\address[label2]{Mani L. Bhaumik Institute for Theoretical Physics, University of California, Los Angeles, California 90095, USA}
\address[label3]{Theoretical Division, Los Alamos National Laboratory, Los Alamos, New Mexico 87545, USA}
\address[label4]{Science Division, Penn State University Berks, Reading, Pennsylvania 19610, USA}
\address[label5]{Jefferson Lab, 12000 Jefferson Avenue, Newport News, Virginia 23606, USA}
\address[label6]{Nuclear Science Division, Lawrence Berkeley National Laboratory, Berkeley, California 94720, USA}

\begin{abstract}
We investigate the Collins azimuthal asymmetry of hadrons produced inside jets in transversely polarized proton-proton collisions. Recently, the quark transversity distributions and the Collins fragmentation functions have been extracted within global analyses from data of the processes semi-inclusive deep inelastic scattering and electron-positron annihilation. We calculate the Collins azimuthal asymmetry for charged pions inside jets using these extractions for RHIC kinematics at center-of-mass energies of 200 and 500 GeV. We compare our results with recent data from the STAR Collaboration at RHIC and find good agreement, which confirms the universality of the Collins fragmentation functions. In addition, we further explore the impact of transverse momentum dependent evolution effects. 
\end{abstract}


\end{frontmatter}

\section{Introduction}
The transverse momentum dependent parton distribution functions and fragmentation functions have recently received an increased interest from both the experimental and theoretical communities~\cite{Accardi:2012qut,Boer:2011fh,Aschenauer:2014twa}. Transverse momentum dependent distributions (TMDs) provide new information about the nucleon structure, in particular for the three-dimensional imaging of the nucleon in momentum space. At the same time, TMDs open new windows for a better understanding of the most fundamental and interesting aspects of QCD, such as gauge invariance and universality properties. 

One of the widely discussed TMDs is the Collins fragmentation function~\cite{Collins:1992kk}. It describes a transversely polarized quark fragmenting into an unpolarized hadron. The hadron's transverse momentum with respect to the direction of the fragmenting quark correlates with the transverse polarization vector of the quark. The Collins fragmentation functions generate azimuthal angular asymmetries in the production of hadrons in high energy scattering processes. For example, in semi-inclusive deep inelastic scattering (SIDIS) of leptons on the transversely polarized nucleons, an azimuthal single transverse spin asymmetry has been observed by several collaborations including the HERMES Collaboration~\cite{Airapetian:2004tw,Airapetian:2010ds}, the COMPASS Collaboration~\cite{Adolph:2012sn}, and the JLab HALL A experiment~\cite{Qian:2011py}. Such an azimuthal correlation is usually referred to as the Collins asymmetry. The modulation is proportional to $\sin(\phi_s+\phi_h)$, where $\phi_s$ and $\phi_h$ are the azimuthal angles of the transverse spin of the nucleon and of the final-state hadron's transverse momentum, respectively. The asymmetry is generated through the quark transversity distributions in the nucleon~\cite{Ralston:1979ys,Jaffe:1991kp,Jaffe:1991ra} coupled with the Collins fragmentation functions. 

The Collins fragmentation functions can also contribute to an azimuthal angular correlation in back-to-back hadron production in electron-positron annihilation~\cite{Boer:1997mf}. In this case, the correlation has a $\cos(2\phi)$ modulation, where $\phi$ is the azimuthal angle between the two hadrons. It is generated through the convolution of two Collins fragmentation functions for the observed hadron pair. The resulting $\cos(2\phi)$ azimuthal correlation has now been measured at several facilities. The BELLE and {\em BABAR} Collaborations published data sets taken at the B-factories at a center-of-mass (CM) energy of $\sqrt{s}\simeq 10.6$ GeV~\cite{Abe:2005zx,Seidl:2008xc,TheBABAR:2013yha}, and the BESIII Collaboration performed measurements at the BEPC facility at a CM energy of $\sqrt{s} = 3.65$ GeV~\cite{Ablikim:2015pta}. The combined analyses of the experimental data on SIDIS and electron-positron annihilation have provided important information on both the quark transversity distributions and the Collins fragmentation functions~\cite{Kang:2015msa,Anselmino:2015sxa}. Transversity distributions give information on the quark contributions to the nucleon tensor charge which is a fundamental property of the nucleon.

Another important process to explore the Collins fragmentation functions is to study the azimuthal asymmetries of hadron production inside highly energetic jets in transversely polarized proton-proton collisions at RHIC~\cite{Yuan:2007nd,DAlesio:2010sag}: $p^\uparrow p\to {\rm jet}(\eta, p_T) \, h(z_h, j_\perp)+X$. Here, $\eta$ and $p_T$ are the rapidity and the transverse momentum of the jet measured in the $pp$ CM frame, respectively. Furthermore, $z_h$ is the momentum fraction of the fragmenting quark jet carried by the hadron, and $j_\perp$ is the hadron transverse momentum with respect to the standard jet axis. The $j_\perp$-distribution of hadrons produced inside jets in unpolarized $pp$ collisions was studied recently in~\cite{Bain:2016rrv,Kang:2017glf}. In the transversely polarized $p^\uparrow p$ scatterings, due to the transverse spin transfer in the hard partonic processes~\cite{Stratmann:1992gu,Collins:1993kq}, the final state quark jet inherits the transverse polarization of quarks in the incoming transversely polarized nucleon. Eventually, the transverse spin of the fragmenting quark correlates with the transverse momentum of the hadron with respect to the jet axis, which leads to a nontrivial Collins asymmetry of the azimuthal angular distribution of hadrons inside (quark) jets~\cite{Yuan:2007nd,DAlesio:2010sag}. The detailed study of these azimuthal asymmetries of hadrons produced inside jets is the main focus of this work. We will focus on two main aspects which are the universality and the evolution of the Collins TMD fragmentation functions.

Concerning the universality aspect, it has been shown in~\cite{Yuan:2007nd} that the Collins fragmentation functions are universal in the sense that they are the same for hadrons inside jets as for SIDIS and electron-positron annihilation. This assessment is expected to be true even when the soft factor is included in order to obtain the full TMD factorization formalism~\cite{Kang:2017glf}. Therefore, it is possible to predict the Collins azimuthal asymmetries for hadron production inside jets at RHIC by using the quark transversity distributions and Collins fragmentation functions determined from SIDIS and electron-positron annihilation. The comparison of our results with the experimental data from RHIC~\cite{Adamczyk:2017wld,Adkins:2016uxv,Aschenauer:2016our,Aschenauer:2015eha} then provides an important test of the universality of the Collins fragmentation functions for these different processes. 

The second important aspect of the Collins fragmentation function is its TMD evolution, i.e. the appropriate QCD evolution of TMD sensitive observables~\cite{Collins:2011zzd}. It is crucial to take into account TMD evolution of the Collins fragmentation functions for phenomenological studies since the experimental measurements are usually performed at different scales. The hard momentum scale $Q^2$ ranges between a couple of GeV$^2$ up to several hundred GeV$^2$. In a recent study, the evolution effects have been implemented in a global analysis of the Collins azimuthal asymmetries in SIDIS and electron-positron annihilation~\cite{Kang:2015msa}. It has been demonstrated in~\cite{Kang:2017glf} that the same TMD evolution applies to the relevant TMD fragmentation functions encountered in the transverse momentum distribution of hadrons inside jets. In this work, we assess the impact of TMD evolution for the azimuthal asymmetries for hadrons inside jets by comparing to the available data from RHIC.

The hadron distribution inside fully reconstructed jets has received broad interest from the high energy nuclear and the particle physics communities in the past years~\cite{Aad:2011sc,Chatrchyan:2012gw,Aaij:2017fak,Aaboud:2017tke}. In particular, the transverse momentum distribution of hadrons relative to a predetermined jet axis may provide important new information about the hadronization of particles at current collider experiments. In~\cite{Bain:2016rrv,Kang:2017glf} the standard jet axis was discussed, whereas in~\cite{Neill:2016vbi} a recoil-free axis, e.g. the winner-take-all axis, was considered. Interestingly, the choice of the axis probes different physics of the hadronization process. In this work, we consider the standard jet axis which allows for a direct relation to standard TMDs extracted from other processes. Needless to say that many studies have been carried out where the longitudinal momentum distribution of different hadrons and even photons inside jets were considered~\cite{Procura:2009vm,Jain:2011xz,Arleo:2013tya,Kaufmann:2015hma,Chien:2015ctp,Kang:2016mcy,Dai:2016hzf,Kaufmann:2016nux,Kang:2017yde}. Studying the correlations of hadrons inside jets constitutes a new opportunity to study TMDs besides the traditional observables. In addition, it may shed new light on other interesting topics such as the effect of non-global logarithms as advocated in~\cite{Dasgupta:2001sh,Banfi:2002hw}.

The azimuthal distributions of hadrons inside a jet were also considered in the framework of the so-called Generalized Parton Model (GPM) in Refs.~\cite{DAlesio:2010sag,DAlesio:2013cfy}. In the GPM, one naively uses TMDs for both parton distribution functions and fragmentation functions, and at the same time assumes that all functions are universal. Both statements lack full justification in QCD, and therefore, the GPM cannot include QCD evolution for these functions properly. On the other hand, the factorization formula used in the current paper involves a mixture of collinear and TMD factorization, following~\cite{Yuan:2007nd, Kang:2017glf}. In this formula, the production of the jet involves only a collinear factorization, in which collinear parton distribution functions are used. At the same time, the internal structure of the jet, i.e., the hadron $j_\perp$ distribution inside the jet is given by the TMD factorization, where the proper evolution of the Collins fragmentation function can be studied. This approach is applicable in the narrow jet approximation, i.e. up to corrections that are power suppressed by ${\cal O}(R^2)$. See the next section for a more detailed discussion.

The remainder of this paper is organized as follows. In Sec.~\ref{sec:formalism}, we review the leading order calculation of the Collins azimuthal asymmetry for hadron production inside jets in $pp$ collisions. In addition, we outline how the TMD evolution effects are implemented. We also present the parton model results where no TMD evolution is taken into account. Numerical results are presented in Sec.~\ref{sec:pheno}, by making use of the recent global extractions of the quark transversity distributions and the Collins fragmentation functions. We calculate the Collins azimuthal asymmetry for charged pion production inside jets in proton-proton collisions for both CM energies 200 and 500 GeV. We compare our results with the experimental data from the STAR Collaboration at RHIC and we conclude our paper in Sec.~\ref{sec:summary}.

\section{Theoretical framework}
\label{sec:formalism}
We consider the hadron azimuthal distribution inside jets in transversely polarized $p^\uparrow p$ collisions,
\bea
p^\uparrow(P_A, S_T, \phi_S) + p(P_B) \to {\rm jet} (\eta, p_T) \, h(z_h, j_\perp, \phi_H) + X\,.
\nonumber
\eea
The momentum of the incoming transversely polarized proton is denoted by $P_A$ (moving in the ``$+z$'' direction) and its transverse polarization vector is $S_T$. 
%%%%%%%%%%%%%%%%%%
\bef
\includegraphics[width=0.45\columnwidth]{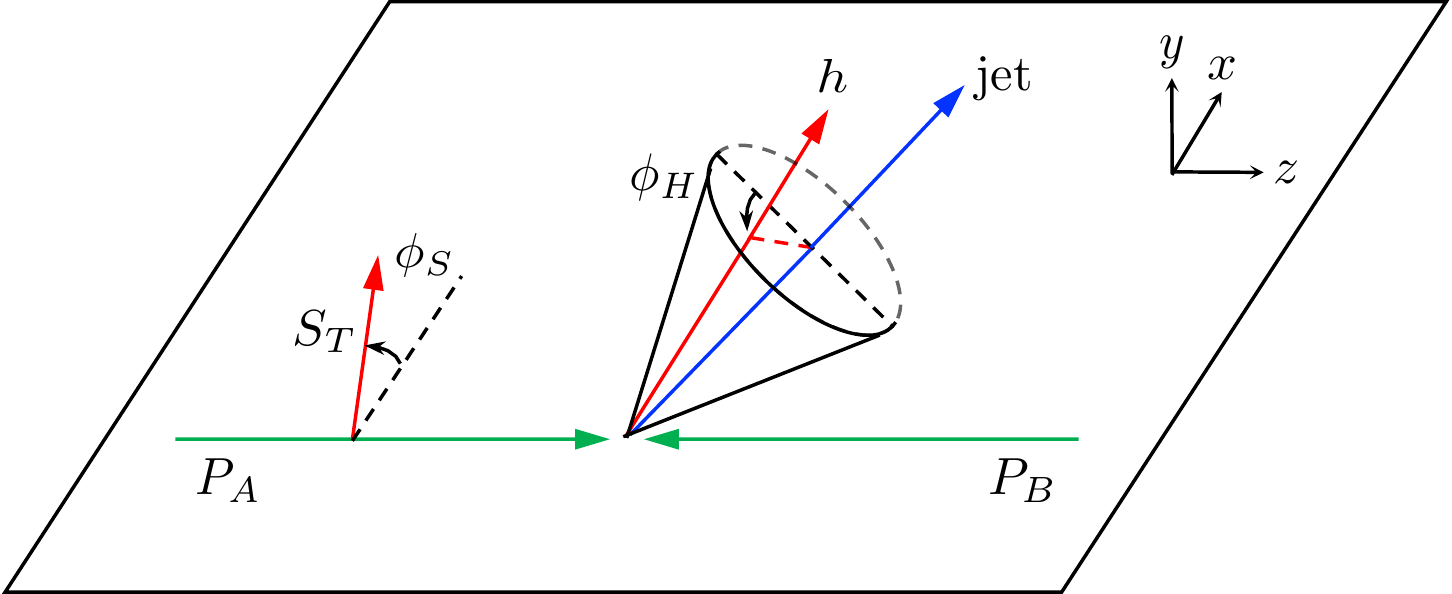}
\caption{Illustration of the relevant kinematic variables for the azimuthal angular distribution of hadrons inside jets in transversely polarized $p^\uparrow p$ collisions. The incident polarized proton has momentum $P_A$ and its transverse polarization vector is denoted by $S_T$. The unpolarized proton has momentum $P_B$. The transverse momentum of the hadron inside the jet relative to the (standard) jet axis is denoted by $j_\perp$.  The azimuthal angles of $S_T$ and $j_\perp$ are defined with respect to the reaction plane and are denoted by  $\phi_S$ and $\phi_H$, respectively.}
\label{fig:illustration}
\eef
%%%%%%%%%%%%%%%%%%
The reaction plane is defined by the two incoming protons and the axis of the observed jet in the final state. We denote the azimuthal angle of the transverse polarization vector $S_T$ with respect to the reaction plane by $\phi_S$. The unpolarized proton (moving in the ``$-z$'' direction) has momentum $P_B$. Moreover, $\eta$ and $p_T$ are the rapidity and transverse momentum of the final state jet. The observed hadron inside that jet is characterized by the following variables: the longitudinal momentum fraction of the jet carried by the hadron is denoted by  $z_h$ and its transverse momentum with respect to the (standard) jet axis is given by $j_\perp$. The hadron transverse momentum vector $j_\perp$ forms an angle $\phi_H$ with the reaction plane. See Fig.~\ref{fig:illustration} for an illustration of the setup of this process and the definition of all the relevant kinematic variables.

\subsection{QCD formalism}

The differential cross section of the hadron azimuthal distribution inside jets can be written as~\cite{Yuan:2007nd}
\bea
\frac{d\sigma}{d\eta d^2p_T dz_h d^2j_\perp} = F_{UU} + \sin(\phi_S - \phi_H) F_{UT}^{\sin(\phi_S - \phi_H)}\,, 
\label{eq:cross_section}
\eea
where $F_{UU}$ and $F_{UT}^{\sin(\phi_S - \phi_H)}$ are the spin-averaged and spin-dependent structure functions, respectively. The so-called Collins azimuthal spin asymmetry $A_{UT}^{\sin(\phi_S - \phi_H)}$ is given by the ratio
\bea
\label{eq:AUT}
A_{UT}^{\sin(\phi_S - \phi_H)}(z_h, j_\perp; \eta, p_T) = \frac{F_{UT}^{\sin(\phi_S - \phi_H)}}{F_{UU}}\,.
\eea
The structure functions $F_{UU}$ and $F_{UT}^{\sin(\phi_S - \phi_H)}$ depend on $\eta,~p_T,~z_h$, and $j_\perp$. In the following we will suppress the arguments $\eta$ and $p_T$ and keep only the $z_h$ and $j_\perp$ dependence for simplicity. 

Using QCD factorization at leading order (LO), the structure functions $F_{UU}$ and $F_{UT}^{\sin(\phi_S - \phi_H)}$ can be written as~\cite{Yuan:2007nd,Kang:2017glf}~\footnote{The next-to-leading order (NLO) formalism for $F_{UU}$ was derived in~\cite{Kang:2017glf}. Since there is no corresponding NLO calculations available for $F_{UT}^{\sin(\phi_S - \phi_H)}$, we use the LO hard factors for both $F_{UU}$ and $F_{UT}^{\sin(\phi_S - \phi_H)}$ in our study.}
\bea
\label{eq:FUU}
F_{UU}(z_h, j_\perp) & = \frac{\alpha_s^2}{s} \sum_{a,b,c} \int_{x_{1\rm min}}^1 \frac{dx_1}{x_1} f_{a/A}(x_1, \mu) 
\int_{x_{2\rm min}}^1\frac{dx_2}{x_2} f_{b/B}(x_2, \mu) D_{h/c}(z_h, j_\perp^2; Q) 
H_{ab\to c}^{\rm U}(\hat s, \hat t, \hat u) \delta(\hat s + \hat t + \hat u)\,,
\\
F_{UT}^{\sin(\phi_S - \phi_H)}(z_h, j_\perp) & = \frac{\alpha_s^2}{s} \sum_{a,b,c} \int_{x_{1\rm min}}^1 \frac{dx_1}{x_1} h_1^{a}(x_1, \mu) 
 \int_{x_{2\rm min}}^1\frac{dx_2}{x_2} f_{b/B}(x_2, \mu) \frac{j_\perp}{z_h M_h}H_{1\,h/c}^{\perp}(z_h, j_\perp^2; Q) 
H_{ab\to c}^{\rm Collins}(\hat s, \hat t, \hat u) \delta(\hat s + \hat t + \hat u)\,,
\label{eq:FUT}
\eea
where we sum over all relevant partonic channels $ab\to c$. The CM energy squared is given by $s=(P_A+P_B)^2$, and $M_h$ is the mass of the observed hadron inside the jet. Furthermore, $\alpha_s$ is the strong coupling constant and $\hat s,~\hat t,~\hat u$ are the standard partonic Mandelstam variables. The unpolarized collinear parton distribution functions (PDFs) are denoted by $f_{a/A}(x_1, \mu)$ and $f_{b/B}(x_2, \mu)$, whereas $h_1^{a}(x_1, \mu)$ are the collinear quark transversity distributions in a transversely polarized proton. The evolution of the collinear PDFs with the factorization scale $\mu$ follows the usual DGLAP equations and similarly for the quark transversity distributions $h_1^{a}(x_1, \mu)$~\cite{Barone:1997fh,Hayashigaki:1997dn,Vogelsang:1997ak}. The lower integration limits $x_{1\rm min}$ and $x_{2\rm min}$ are given by
\bea
x_{1\rm min} = \frac{x_Te^\eta}{2-x_T e^{-\eta}}\,, 
\qquad
x_{2\rm min} = \frac{x_1 x_T e^{-\eta}}{2x_1 - x_T e^{\eta}}\,,
\eea
where $x_T = 2p_T/\sqrt{s}$. Moreover, $D_{h/c}(z_h, j_\perp^2; Q)$ are the unpolarized TMD fragmentation functions, and $H_{1\,h/c}^{\perp}(z_h, j_\perp^2; Q)$ are the Collins fragmentation functions in the so-called Trento convention~\cite{Bacchetta:2004jz}. Note that at LO, the hadron transverse momentum with respect to the fragmenting parent quark is equal to $j_\perp$, i.e. the hadron transverse momentum with respect to the jet axis. Going beyond LO, they are no longer equal to each other due to the effect of soft gluon radiation but they are still closely related, see \cite{Kang:2017glf} for more details. The momentum scale $Q$ represents the appropriate factorization scale for both the unpolarized TMD fragmentation functions and the Collins fragmentation functions~\cite{Collins:2011zzd}. The $Q$-dependence of these TMD functions is generally referred to as TMD evolution. Here, we use $Q$ to emphasize that the TMD evolution is different from the DGLAP evolution of the collinear PDFs associated with the scale $\mu$. We will discuss this aspect in more detail below, which has been studied extensively in the literature~\cite{Kang:2015msa,Aybat:2011zv,Sun:2013hua,Echevarria:2014xaa,Echevarria:2014rua,Echevarria:2012pw,Aybat:2011ge,Kang:2011mr}. The hard functions $H_{ab\to c}^{\rm U}$ for the unpolarized structure function are well-known, and are available for example in Ref.~\cite{Owens:1986mp}. The corresponding hard functions $H_{ab\to c}^{\rm Collins}$ for the spin-dependent structure function are also available in the literature~\cite{Yuan:2007nd}. They are exactly the same as those calculated for the transverse spin transfer in the hard partonic processes~\cite{Stratmann:1992gu,Collins:1993kq}. 

It might be instructive to comment on the factorization formula given in Eqs.~\eqref{eq:FUU} and \eqref{eq:FUT}, which was first written down in \cite{Yuan:2007nd} at LO, and then in \cite{Kang:2017glf} at NLO. As we have emphasized already in the Introduction, such 
a factorized form is a mixture of collinear and TMD factorization, involving two steps: the first step is a collinear factorization for the production of the jet, involving collinear PDFs; while the second step is a TMD factorization for the hadron $j_\perp$-distribution inside the jet. The factorization arguments were provided in \cite{Kang:2017glf} within the {\it standard} soft-collinear effective theory (SCET)~\cite{Bauer:2000ew,Bauer:2000yr,Bauer:2001ct,Bauer:2001yt,Bauer:2002nz}. In other words, the issue of the spectator interactions found by Collins and Qiu~\cite{Collins:2007nk} (represented by ``Glauber modes'' in SCET) is not considered, and it deserves a further investigation along the lines of \cite{Rothstein:2016bsq}. 

In the remainder of this section, we are going to discuss the details of the unpolarized TMD fragmentation functions and the Collins fragmentation functions. In the next section, we first provide the results for these TMDs including the full TMD evolution. For comparison, we also present parton model results where no TMD evolution is taken into account. In this case, we choose a simple Gaussian form for the transverse momentum dependence of both TMDs.

\subsection{Results with TMD evolution}
It has been shown that the Collins fragmentation function $H_{1\,h/c}^{\perp}$ is universal in different processes~\cite{Yuan:2007nd,Metz:2002iz,Collins:2004nx,Meissner:2008yf,Gamberg:2008yt}, including SIDIS, $e^+e^-$ annihilation, and the process studied
in this paper. The key observation in these studies is that the 
eikonal propagators do not generate the phases necessary for 
a non-zero Collins asymmetries in these processes. This
has been demonstrated explicitly at the two-gluon exchange order for hadron distribution inside the jet~\cite{Yuan:2008yv}. 
We expect the same conclusion 
holds at even higher orders. Because of this property, we can apply the Ward identity to sum all initial- and final-state interaction effects into the gauge link associated with the
TMD fragmentation function. 
This is very different from that for
the TMD parton distributions, where the eikonal propagators
contribute to a non-zero phase and lead to the non-universality
of the so-called Sivers functions among different processes. 
In particular, much more complicated results have been found for 
the single spin asymmetries in the hadronic dijet correlations where a normal TMD 
factorization breaks down~\cite{Collins:2007nk,Vogelsang:2007jk,Collins:2007jp,Rogers:2010dm}. 
The reason is precisely that the eikonal propagators from the initial- and final-state interactions in 
dijet correlation process do contribute to the poles in the cross 
section~\cite{Rogers:2010dm}. Because of this, the Ward identity is not applicable, 
and the standard TMD factorization breaks down. 

The modern and proper definition of TMDs usually includes the soft factor for the specific process, which captures the contribution from soft gluon radiation, see Ref.~\cite{Collins:2011zzd} for more details. Within the standard SCET framework, it was further demonstrated in~\cite{Kang:2017glf} that when the soft factor is included, the combination of the TMD fragmentation function and the soft function as probed inside jets in $pp$ collisions is the same as in SIDIS and electron-positron annihilation. Of course, it would be desirable to revisit this conclusion when including the Glauber modes in SCET, as mentioned in last section. At the same time, investigating the initial- and final-state interactions beyond the two-gluon exchange could also be very useful. For the rest of this section, we will use the same TMD evolution for calculating the distribution of hadrons inside jets.

In fact it was demonstrated~\cite{Kang:2017glf} that the relevant scale $Q$ for the TMD fragmentation functions for the hadron distribution inside jets is given by the natural scale set by the jet dynamics which is $p_T R$, where $R$ is the jet size parameter. With this choice, one may further perform an additional evolution from the scale $p_T R$ to $p_T$ in order to resum single logarithms in the jet size parameter to all orders in the strong coupling constant $\alpha_s^n \ln^n R$. For our current study, we only work to LO in QCD which is independent of the jet size parameter $R$. We will thus use $Q=p_T$ for the TMDs in Eqs.~\eqref{eq:FUU} and \eqref{eq:FUT}, i.e. $D_{h/c}(z_h, j_\perp^2; Q = p_T)$ and $H_{1\,h/c}^{\perp}(z_h, j_\perp^2; Q=p_T)$, respectively. For completeness, we note that we also use $\mu=p_T$ for the collinear PDFs and the quark transversity distributions in Eqs.~\eqref{eq:FUU} and \eqref{eq:FUT}.

A global fit of both the quark transversity distributions and the Collins fragmentation functions from SIDIS and electron-positron annihilation data was performed recently in~\cite{Kang:2015msa}, where the effects of TMD evolution were studied in detail. Using these extracted functions, we can then calculate the Collins azimuthal asymmetries for the hadron distribution inside jets in $p^\uparrow p$ collisions, and compare with the recent experimental measurements from RHIC. We now provide a short review of the TMD evolution as it is used in this work where we closely follow the results presented in Ref.~\cite{Kang:2015msa}. The two TMD fragmentation functions can be written as 
%\begin{widetext}
\begin{subequations}
\label{eq:D1_tmd}
\bea
D_{h/q}(z_h, j_\perp^2; Q) =& \frac{1}{z_h^2}\int_0^{\infty} \frac{db\; b}{(2\pi)} J_0(j_\perp b/z_h)\, \hat C_{i\gets q}^{D_1}\otimes D_{h/i}(z_h, \mu_b) \, e^{-\frac{1}{2} S_{\rm pert}(Q, b_*)  - S_{\rm NP}^{D_1}(Q, b)}\,, 
\\
\frac{j_\perp}{z_h M_h} H_{1\,h/q}^\perp(z_h, j_\perp^2; Q)  = & \frac{1}{z_h^2}\int_0^{\infty} \frac{db\; b^2}{(2\pi)} J_1(j_\perp b/z_h)
\, \delta \hat C_{i\gets q}^{\rm collins} \otimes \hat H^{\perp(1)}_{1\,h/i}(z_h, \mu_b)\,
e^{-\frac{1}{2} S_{\rm pert}(Q, b_*)  - S_{\rm NP}^{\rm collins}(Q, b)}\,.
\eea
\end{subequations}
%\end{widetext}
Here, we denote $b=|{\boldsymbol b}|$, where ${\boldsymbol b}$ is the 2-dimensional coordinate variable conjugate to the transverse momentum component $j_\perp$. In addition, we have $\mu_b=c_0/b$ with $c_0 = 2e^{-\gamma_E}$, and $\otimes$ represents a convolution in the momentum fraction $z_h$, e.g.
\bea\label{eq:TMDresult}
\hat C_{i\gets q}^{D_1}\otimes D_{h/i}(z_h, \mu_b) 
= \sum_{i} \int_{z_h}^1\frac{dz_h'}{z_h'} \hat C_{i\gets q}^{D_1}\left(\frac{z_h}{z_h'}, \mu_b\right) D_{h/i}(z_h', \mu_b)\,,
\eea
where $D_{h/i}(z_h, \mu_b)$ are the usual collinear fragmentation functions. The collinear twist-3 functions $\hat H^{\perp(1)}_{1\,h/i}(z_h, \mu_b)$ are equal to the first moment of the Collins fragmentation functions. The functional form of $\hat H^{\perp(1)}_{1\,h/i}(z_h, \mu)$ was determined by means of a global analysis of the Collins asymmetry in SIDIS and electron-positron annihilation, see Ref.~\cite{Kang:2015msa} for details. The coefficient functions for the unpolarized and polarized case are denoted by $\hat C_{i\gets q}^{D_1}$ and $\delta \hat C_{i\gets q}^{\rm collins}$, respectively. Their expressions up to the next-to-leading order (NLO) are given by~\cite{Kang:2015msa} 
\bea
\hat C_{q'\gets q}(z_h,\mu_b) = &\delta_{q'q} \bigg[\delta(1-z_h)+\frac{\alpha_s}{\pi}\Big(\frac{C_F}{2}(1-z_h) 
 + P_{q\gets q}(z_h)\, \ln z_h \Big) \bigg]\; ,  \label{eq:cd}
\\
\hat C_{g\gets q}(z_h,\mu_b) =& \frac{\alpha_s}{\pi} \left( \frac{C_F}{2} z_h\; +  P_{g\gets q}(z_h)\, \ln z_h \right) ,  \label{eq:cd1}\\
\delta \hat C_{q'\gets q}^{\rm collins}(z_h,\mu_b) =&\delta_{q'q} \left[\delta(1-z_h)+\frac{\alpha_s}{\pi}\hat P_{q\gets q}^{c}(z_h) \ln z_h \right]\; 
\label{eq:ch1perp},
\eea
with the relevant splitting functions given by
\bea
P_{q\gets q}(z_h) &= C_F \left[ \frac{1+z_h^2}{(1-z_h)}_+ + \frac{3}{2} \delta(1-z_h) \right] \, , 
\label{P_qq}
\\
P_{g\gets q}(z_h) &= C_F \frac{1+(1-z_h)^2}{z_h} \; ,\\
\hat P_{q\gets q}^{c}(z_h) &= C_F \left[ \frac{2 z_h}{(1- z_h)}_+ +\frac{3}{2}\delta(1-z_h)\right].
\label{P_gq}
\eea
Note that the coefficient functions for the unpolarized TMD functions are available up to the next-to-next-to leading order~\cite{Echevarria:2016scs}. 

The perturbative Sudakov factor in Eq.~\eqref{eq:TMDresult} can be written as
\bea
S_{\rm pert}(Q,b)=\int_{\mu_b}^{Q}\frac{d\mu'}{\mu'}\left[A\,\ln\left(\frac{Q^2}{\mu'^2}\right)+B\right] \ ,
\eea
where the coefficients $A$ and $B$ can be calculated perturbatively as $A=\sum_{n=1}A^{(n)}(\alpha_s/\pi)^n$ and $B=\sum_{n=1}B^{(n)}(\alpha_s/\pi)^n$. For our phenomenological results, we work at NLL accuracy, and we thus take into account the coefficients $A^{(1)},~A^{(2)},~B^{(1)}$. For completeness, we list the relevant results here~\cite{Kang:2011mr,Aybat:2011zv,Echevarria:2012pw,Collins:1984kg,Qiu:2000ga,Landry:2002ix}
\bea
A^{(1)}&=C_F\,, 
\\
A^{(2)}& = \frac{C_F}{2}\left[C_A\left(\frac{67}{18} - \frac{\pi^2}{6}\right) - \frac{10}{9} T_F n_f \right]\,,
\\
B^{(1)}& =-\frac{3}{2}C_F\,. 
\eea
It is well-known that the TMD evolution contains a non-perturbative piece in the region where $1/b\gg \Lambda_{\rm QCD}$. This is why one has to introduce a prescription to extrapolate between the perturbative small-$b$ region and the non-perturbative large-$b$ region. In this work, we choose to adopt the standard $b_*$-prescription~\cite{Collins:1984kg}. Alternative approaches can be found in~\cite{Kulesza:2002rh,Qiu:2000hf,Catani:2015vma,Ebert:2016gcn,Monni:2016ktx}. One defines $b_*$ as
\bea
b_* = \frac{b}{\sqrt{1+b^2/b_{\rm max}^2}}\;,
\eea
such that $b_*\to b_{\rm max}$ for large $b$. Here, $b_{\rm max}$ is a parameter of the prescription, which was chosen as $b_{\rm max} = 1.5$ GeV$^{-1}$ in the global analysis of~\cite{Kang:2015msa}. After introducing $b_*$ in the Sudakov factor, the total Sudakov factor includes a non-perturbative contribution, besides the perturbative piece $S_{\rm pert}(Q,b_*)$. The non-perturbative Sudakov factors $S_{\rm NP}(Q, b)$ for both the unpolarized TMD fragmentation functions and the Collins fragmentation functions are given by
\bea
S_{\rm NP}^{D_1}(Q, b) &=  \frac{g_2}{2} \ln\left(\frac{b}{b_*}\right)\ln\left(\frac{Q}{Q_0}\right) + \frac{g_h}{z_h^2}\, b^2 \; ,
\\
S_{\rm NP}^{\rm collins}(Q, b) &=  \frac{g_2}{2} \ln\left(\frac{b}{b_*}\right)\ln\left(\frac{Q}{Q_0}\right) + \frac{g_h - g_c}{z_h^2}\, b^2 \; .
\label{eq:softfactores}
\eea
We adopt the values $Q_0^2 = 2.4$ GeV$^2$, $g_2 = 0.84$, and $g_h = 0.042$ GeV$^2$ from the analysis of the spin-averaged cross section~\cite{Su:2014wpa}. In addition, we use $g_c = 0.0236\pm 0.0007$ GeV$^2$ following the analysis of the Collins asymmetry~\cite{Kang:2015msa}, as mentioned above. Note that in our parametrization of the non-perturbative Sudakov function is $\ln({b}/{b_*})\propto \ln(1+b^2/b_{\rm max}^2)$ which is crucial for accommodating low-$Q^2$ data in the analysis~\cite{Kang:2015msa}. This parametrization is consistent at small $b$ with the standard $b^2$ parametrization that was used in high-$Q^2$ extractions, for instance that of Ref.~\cite{Konychev:2005iy}.

With all these ingredients at hand, we can then use Eq.~\eqref{eq:D1_tmd} in combination with Eqs.~\eqref{eq:FUU} and \eqref{eq:FUT} to compute the Collins azimuthal asymmetry $A_{UT}^{\sin(\phi_S - \phi_H)}$ for the hadron distribution inside jets in $p^\uparrow p$ collisions including TMD evolution. 

\subsection{Results without TMD evolution}
\label{subsec:withoutTMD}
The quark transversity distributions and the Collins fragmentation functions have also been extracted within a global analysis in the parton model framework in Ref.~\cite{Anselmino:2015sxa}, i.e. without TMD evolution. In this study a Gaussian form of the transverse momentum dependence of the TMD fragmentation functions was adopted. Hence, the unpolarized TMD fragmentation functions and Collins fragmentation functions are written as
\bea\label{eq:FF1}
D_{h/q}(z_h, j_\perp^2; Q) &= D_{h/q}(z_h, Q)\, g(j_\perp)\,,
\\
\frac{j_\perp}{z_hM_h}H_{1\, h/q}^{\perp}(z_h, j_\perp^2; Q) &= {\mathcal N}_q^C(z_h)\, h(j_\perp)\, D_{h/q}(z_h, j_\perp^2; Q)\,.\label{eq:FF2}
\eea
Here, $D_{h/q}(z_h, Q)$ are the standard unpolarized collinear fragmentation functions. The respective $j_\perp$-dependent parts are given by
\bea
g(j_\perp) = \frac{1}{\pi \langle j_\perp^2\rangle} e^{-j_\perp^2/ \langle j_\perp^2\rangle}\,,
\qquad
h(j_\perp) = \sqrt{2e} \frac{j_\perp}{M_C} e^{-j_\perp^2/M_C^2}\,,
\label{eq:h(jT)}
\eea
with $\langle j_\perp^2\rangle = 0.12$ GeV$^2$ which was obtained from an analysis of SIDIS hadron multiplicity data~\cite{Anselmino:2013lza}. The collinear functions ${\mathcal N}_q^C(z_h)$ are parametrized for the so-called favored and disfavored Collins fragmentation functions as
\bea
\label{eq:N1}
{\mathcal N}_{\rm fav}^C(z_h) =N_{\rm fav}^C\, z_h^{\gamma} (1-z_h)^{\delta} \,\frac{(\gamma+\delta)^{\gamma+\delta}}{\gamma^\gamma \delta^\delta}\,,
\qquad
{\mathcal N}_{\rm dis}^C(z_h) &=N_{\rm dis}^C\,. 
\eea
The parameters $M_C$ in Eq.~\eqref{eq:h(jT)}, and $(N_{\rm fav}^C, ~N_{\rm dis}^C, ~\gamma, ~\delta)$ in Eq.~\eqref{eq:N1} were determined within a global analysis of the Collins asymmetry in SIDIS and electron-positron annihilation.

It is instructive to note that within the parton model framework, the $Q$-dependence is only contained in the collinear functions $D_{h/q}(z_h, Q)$, which follow the usual DGLAP evolution equations. In other words, no TMD evolution is considered here. Note that here the Collins fragmentation functions $H_{1\, h/q}^{\perp}(z_h, j_\perp^2; Q)$ in Eq.~\eqref{eq:FF2} are written in terms of the unpolarized TMD fragmentation functions $D_{h/q}(z_h, j_\perp^2; Q)$. Due to the factorization of the $z_h$ and the $j_\perp$ dependence in Eqs.~\eqref{eq:FF1} and~\eqref{eq:FF2}, the shape of the $j_\perp$-dependence of the asymmetry $A_{UT}^{\sin(\phi_S - \phi_H)}$ is directly given by the function $h(j_\perp)$. 

\section{Phenomenology at RHIC}
\label{sec:pheno}
In this section, we present numerical results for the Collins azimuthal asymmetries for hadron production within jets $p^\uparrow p \to ({\rm jet} \, h) + X$, and compare to the experimental measurements by the STAR Collaboration at RHIC. As pointed out in the Introduction and the previous section, the comparison to this data using previously extracted TMDs provides a unique opportunity to test the universality of the involved TMDs and to assess the impact of TMD evolution.

\bef
\includegraphics[width=0.5\columnwidth]{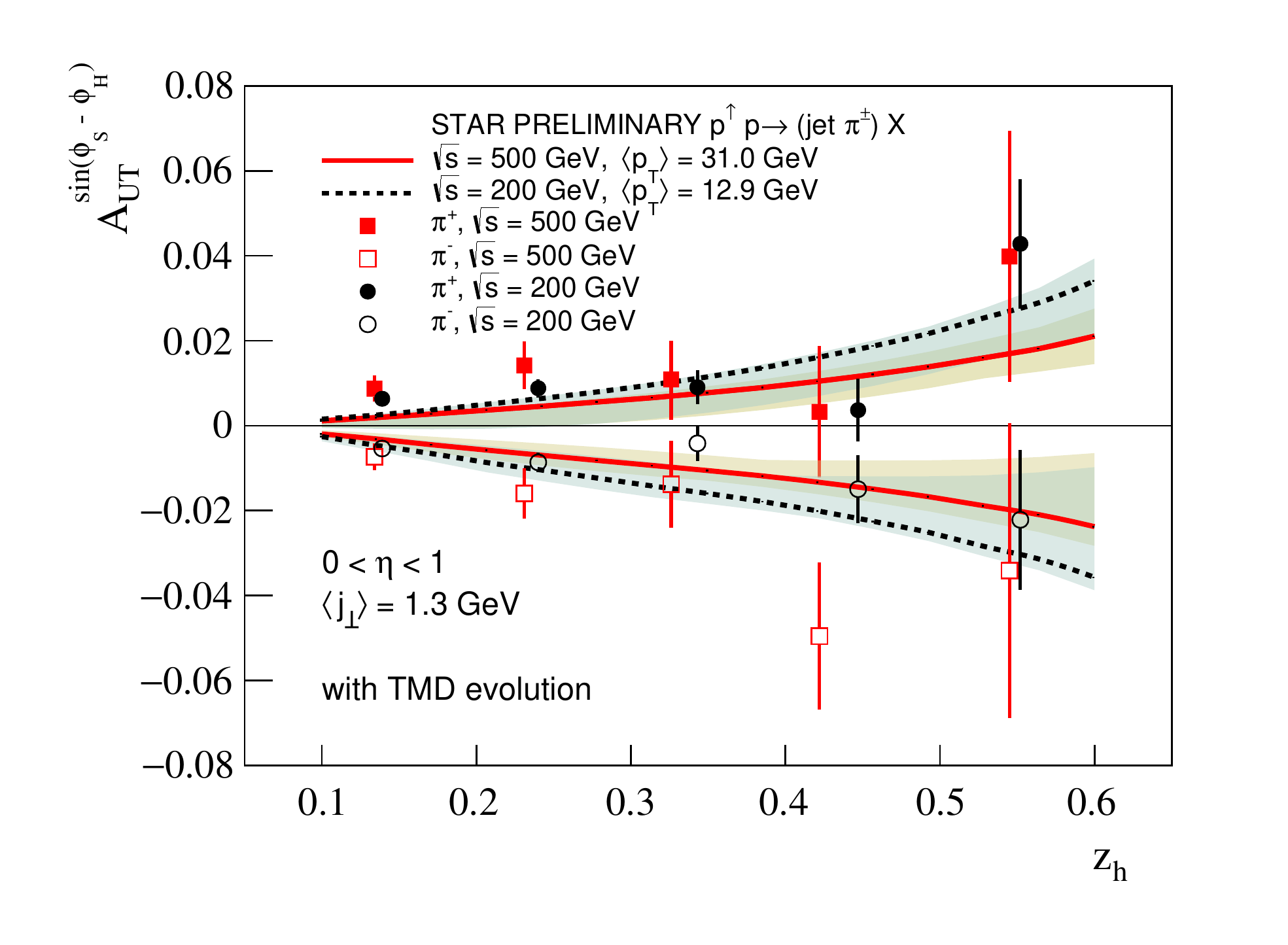}
\caption{The Collins azimuthal spin asymmetry $A_{UT}^{\sin(\phi_S - \phi_H)}$ for pions produced inside jets $pp\to(\text{jet}\,\pi^{\pm})+X$ with TMD evolution using the extracted TMDs of~\cite{Kang:2015msa}. We show our results as a function of $z_h$ compared to the preliminary STAR data of~\cite{Adkins:2016uxv,Aschenauer:2016our,Aschenauer:2015eha} at $\sqrt{s}=200$ GeV (dashed black lines, solid black circles for $\pi^+$, open black circles for $\pi^-$) and $\sqrt{s}=500$ GeV ~\cite{Adamczyk:2017wld} (solid red lines, solid red squares for $\pi^+$, open red squares for $\pi^-$). We have $\langle p_T\rangle = 12.9$ (31.0) GeV for the average jet transverse momentum at $\sqrt{s} = 200$ (500) GeV. The averaged hadron transverse momentum with respect to the (standard) jet axis is given by $\langle j_\perp\rangle = 1.3$ GeV and the jet rapidity is integrated over the range $0<\eta<1$. The error bands are computed using results of Ref.~\cite{Kang:2015msa}.}
\label{fig:withTMD}
\eef

The STAR Collaboration at RHIC has performed measurements of the Collins azimuthal asymmetries $A_{UT}^{\sin(\phi_S - \phi_H)}$ for pion production inside jets in transversely polarized proton-proton collisions $p^\uparrow p \to \left({\rm jet} \, \pi^{\pm}\right) + X$~\cite{Adamczyk:2017wld,Adkins:2016uxv,Aschenauer:2016our,Aschenauer:2015eha}. The jets are reconstructed using the anti-$k_{T}$ algorithm~\cite{Cacciari:2011ma} with a jet size parameter of $R=0.6$~\cite{Adkins:2016uxv}. The measurements were performed separately for charged pions $\pi^+$ and $\pi^-$, and for both CM energies $\sqrt{s} = 200$ and 500 GeV. We use the global extractions of the quark transversity distributions and the Collins fragmentation functions of Refs.~\cite{Kang:2015msa,Anselmino:2015sxa} to perform our numerical calculations. It might be instructive to point out that despite of the wealth of SIDIS data, the kinematic reach of existing SIDIS experiments is still limited to the relatively small Bjorken-$x$ region with $x\lesssim 0.3$. The current and future STAR measurements of jets produced in the forward rapidity region probe the transversity distributions for relatively large values of $x>0.3$. Together with future measurements from Jefferson Lab at 12 GeV, they will provide further constraints~\cite{Ye:2016prn} on the large-$x$ behavior of the transversity distributions. 

Before we present the comparison of our numerical results with the experimental data, we comment on the potential impact of the relevant gluon TMD fragmentation functions. Since there is no gluon transversity distribution nor a gluon Collins fragmentation function, only the respective quark contributions are relevant for the calculation of $F_{UT}^{\sin(\phi_S-\phi_H)}$. However, the unpolarized gluon TMD fragmentation function does contribute to $F_{UU}$. The current STAR experimental data was measured for relatively large jet transverse momenta at forward rapidities and for large values of $z_h$ for charged pions. In this kinematic region, the quark TMD fragmentation functions are expected to dominate. Therefore, we do not include the gluon TMD fragmentation function in our numerical studies.

We start by presenting our numerical results where TMD evolution effects are fully incorporated. For our numerical evaluations, we choose all relevant non-perturbative PDFs and fragmentation functions as in Ref.~\cite{Kang:2015msa}. In Fig.~\ref{fig:withTMD}, we present our results for the Collins azimuthal asymmetry $A_{UT}^{\sin(\phi_S - \phi_H)}$ with TMD evolution as a function of $z_h$. We show the comparison with the data from STAR both at $\sqrt{s}=200$ GeV and $\sqrt{s}=500$ GeV. The jet rapidity is integrated over $0<\eta<1$. The solid red curves are for $\sqrt{s} = 500$ GeV, whereas the dashed black curves are for $\sqrt{s} = 200$ GeV. The presented error bands for our calculations are based on the uncertainties from the quark transversity distributions and the Collins fragmentation functions following Ref.~\cite{Kang:2015msa}. The solid circles (squares) show the experimental data for $\pi^+$ production at $\sqrt{s} = 200$ (500) GeV. The open circles (squares) show the available data for $\pi^-$ production at $\sqrt{s} = 200$ (500) GeV. The averaged jet transverse momentum is given by $\langle p_T\rangle = 12.9$ (31.0) GeV for $\sqrt{s} = 200$ (500) GeV. The values for $\langle p_T\rangle$ were chosen by the experimental collaboration such that for both beam energies roughly the same parton momentum fraction $x$ of the quark transversity distributions is probed. For the experimental data points, the reported averaged hadron transverse momentum with respect to the (standard) jet axis is given by $\langle j_\perp\rangle = 1.3$ GeV. 

\bef
\includegraphics[width=0.5\columnwidth]{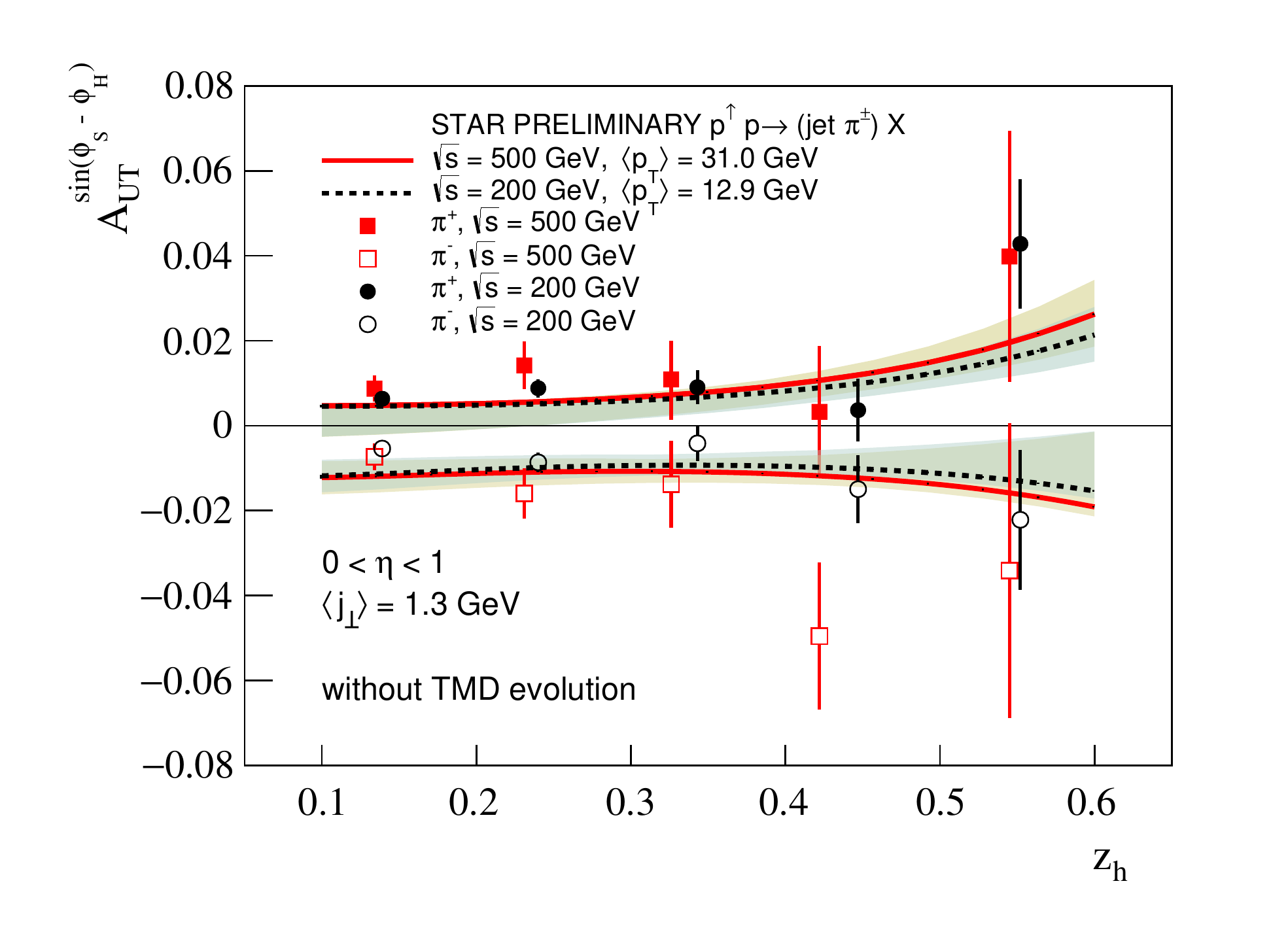}
\caption{Same as Fig.~\ref{fig:withTMD}, but here our results are presented without TMD evolution using the TMDs of Ref.~\cite{Anselmino:2015sxa} as input.}
\label{fig:woutTMD}
\eef

It is evident from Fig.~\ref{fig:withTMD} that our calculations using the quark transversity distributions and the Collins fragmentation functions as extracted from SIDIS and electron-positron annihilation yield a good qualitative description of the experimental data from STAR. The observed agreement within the experimental uncertainties confirms for the first time the universality of the Collins fragmentation functions for the three different processes. Concerning TMD evolution effects, we find that our calculations give slightly smaller Collins azimuthal asymmetries $A_{UT}^{\sin(\phi_S - \phi_H)}$ for the larger jet transverse momentum value $\langle p_T\rangle = 31.0$ GeV compared to the result for $\langle p_T\rangle = 12.9$ GeV. This is consistent with the effect of TMD evolution which typically dilutes the spin asymmetry at a larger momentum scale, as observed for SIDIS in~\cite{Kang:2015msa}. This can be understood as follows. When increasing the momentum scale $Q = p_T$, the $j_\perp$ dependence of the TMDs typically becomes broader, i.e. it is spread out to relatively large values of $j_\perp$. As a result, for fixed $j_\perp$, the relevant TMDs become smaller. Of course, the actual situation is more intricate, as the experimental data for the asymmetry is a ratio of convolutions of TMDs. 

\bef
\includegraphics[width=0.45\columnwidth]{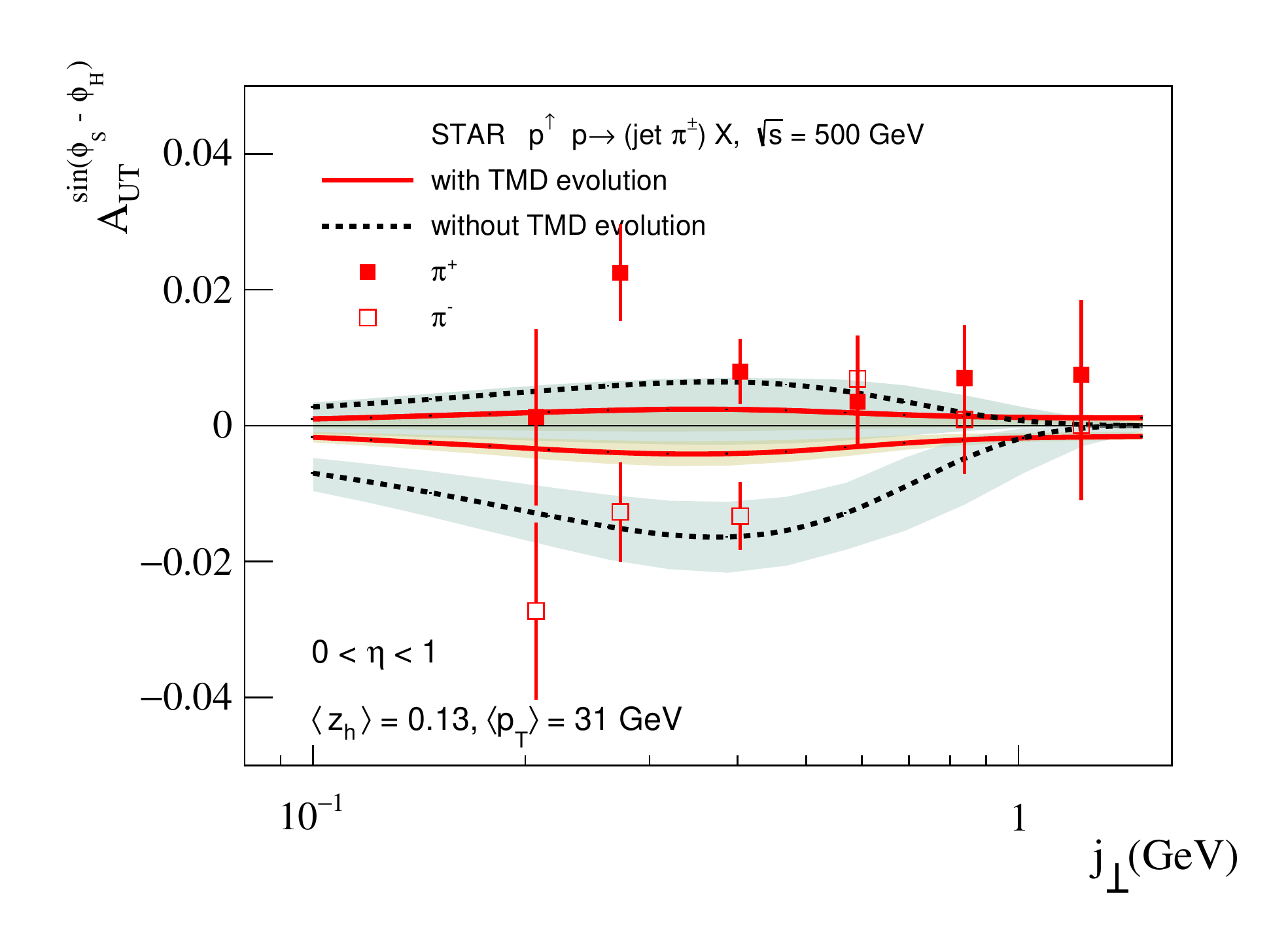}
\includegraphics[width=0.45\columnwidth]{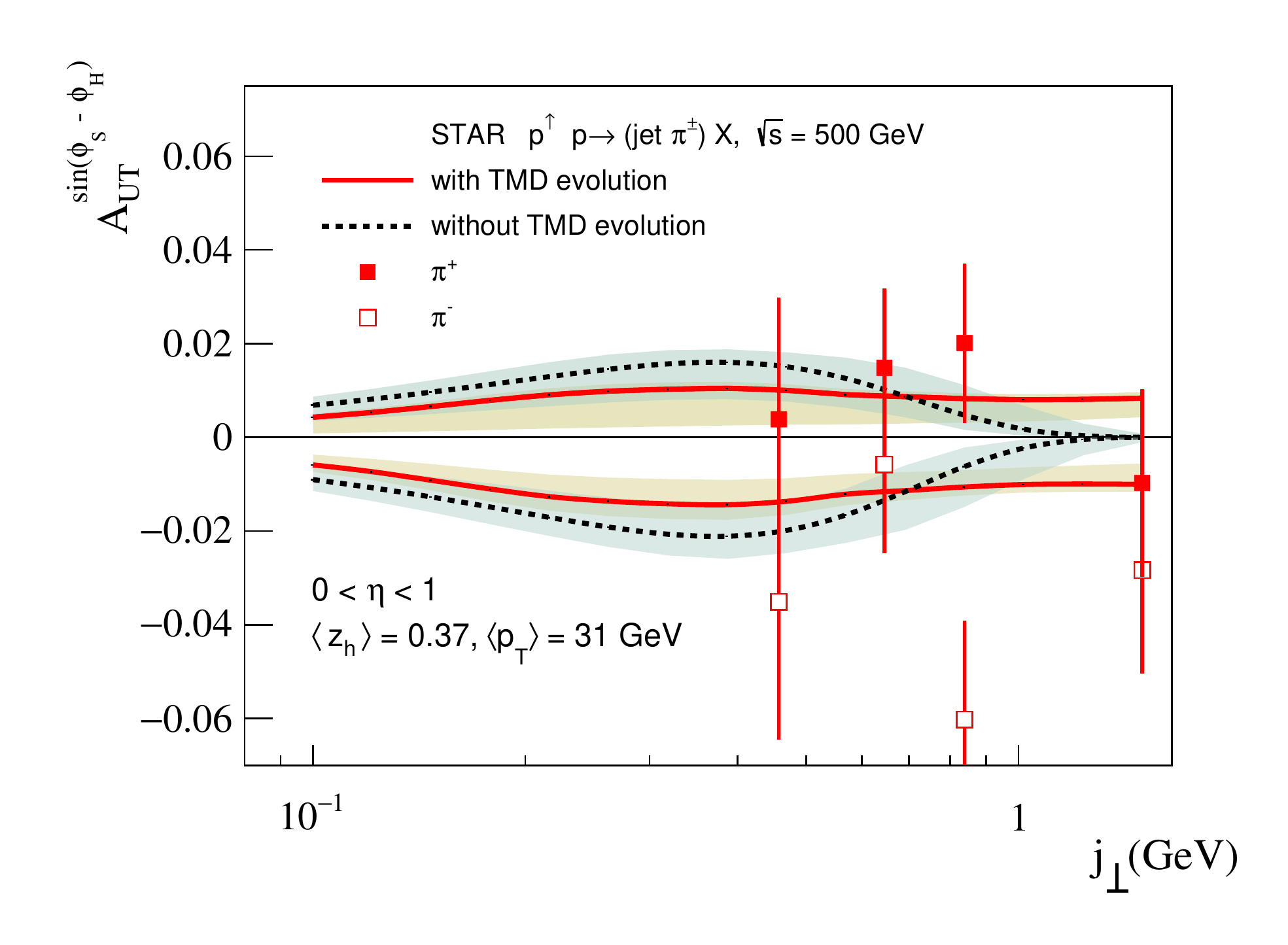}
\caption{The Collins azimuthal spin asymmetry $A_{UT}^{\sin(\phi_S - \phi_H)}$ at $\sqrt{s}=500$ GeV ~\cite{Adamczyk:2017wld} for pions produced inside jets $pp\to(\text{jet}\,\pi^{\pm})+X$ as a function of the pion momentum $j_\perp$ for $\langle z_h\rangle = 0.13$ (left panel) and $\langle z_h\rangle = 0.37$ (right panel). Solid lines correspond to calculations that take into account TMD evolution using the extracted TMDs of Ref.~\cite{Kang:2015msa} and the dashed lines correspond to extraction of TMDs without TMD evolution of Ref.~\cite{Anselmino:2015sxa}. We have $\langle p_T\rangle = 31$  GeV for the average jet transverse momentum and the jet rapidity is integrated over the range $0<\eta<1$. The error bands are computed using results of Refs.~\cite{Kang:2015msa,Anselmino:2015sxa}.}
\label{fig:jt}
\eef

In Fig.~\ref{fig:woutTMD}, we present the comparison of our results without TMD evolution with the same experimental data for $A_{UT}^{\sin(\phi_S - \phi_H)}$. In this case, we use the quark transversity distributions, the unpolarized TMD fragmentation functions and the Collins fragmentation functions of~\cite{Anselmino:2015sxa} based on the parton model framework which does not include the effects of TMD evolution. Again, we obtain a good description of the data within the uncertainties. The presented error bands of our results are again based on the uncertainties of the quark transversity distributions and the Collins fragmentation functions according to Ref.~\cite{Anselmino:2015sxa}. The obtained Collins azimuthal asymmetries $A_{UT}^{\sin(\phi_S-\phi_H)}$ are very similar for both $\sqrt{s}=200$~GeV and 500~GeV. This is due to the absence of TMD evolution effects and one samples the same $x$ and $j_\perp$ values for both CM energies even though the average jet transverse momenta $\langle p_T \rangle$ are quite different. This result is to be expected since the dependence on the momentum scale $Q$ only enters via the corresponding collinear fragmentation functions as discussed in Sec.~\ref{subsec:withoutTMD}. The evolution is relatively mild as it follows the usual DGLAP evolution equations and, hence, the impact on the asymmetry $A_{UT}^{\sin(\phi_S - \phi_H)}$ is small. 

In Fig.~\ref{fig:jt}, we present the comparison of our calculations of  $A_{UT}^{\sin(\phi_S - \phi_H)}$ as a function of the pion momentum $j_\perp$ and the experimental data at $\sqrt{s}=500$ GeV ~\cite{Adamczyk:2017wld} for $\langle z_h\rangle = 0.13$ (left panel) and $\langle z_h\rangle = 0.37$ (right panel). One can observe that calculations done without TMD evolution (dashed lines) have a characteristic behavior: the asymmetry diminishes quickly and become very small at $j_\perp> 1$ GeV.  It happens due to the underlying gaussian behavior of TMD functions with an energy independent width. Calculations made with TMD evolution using as input functions from Ref.~\cite{Kang:2015msa} become broader due to soft gluon radiation, see Eqs.\eqref{eq:softfactores}, such that even at larger values of $j_\perp$ the asymmetry is not diminishing quickly.  As expected the calculations with TMD evolution show suppression of the maximum and broadening of the asymmetry with respect to calculations without TMD evolution.

We observe that the calculated Collins azimuthal asymmetries with and without TMD evolution both give a good description of the experimental data. In other words, the current experimental data cannot resolve the effects of TMD evolution due to their large uncertainties.  

For completeness in Fig.~\ref{fig:unp_jt} we also present our calculations of unpolarized $pp\to(\text{jet}\,\pi^{+})+X$ cross-section, see Eq.~\eqref{eq:cross_section},  at $\sqrt{s}=500$ GeV  as a function of the pion momentum $j_\perp$ for  $\langle z_h\rangle = 0.37$. As expected, calculations without TMD evolution follow simple gaussian shape that is much narrower compared to calculation with TMD evolution. One can see from Fig.~\ref{fig:unp_jt} that effects of evolution are much more dramatic in unpolarized distributions compared to asymmetries. An experimental study of the cross-section as function of  $j_\perp$ will be very helpful in order to study evolution effects.

\bef
\includegraphics[width=0.45\columnwidth]{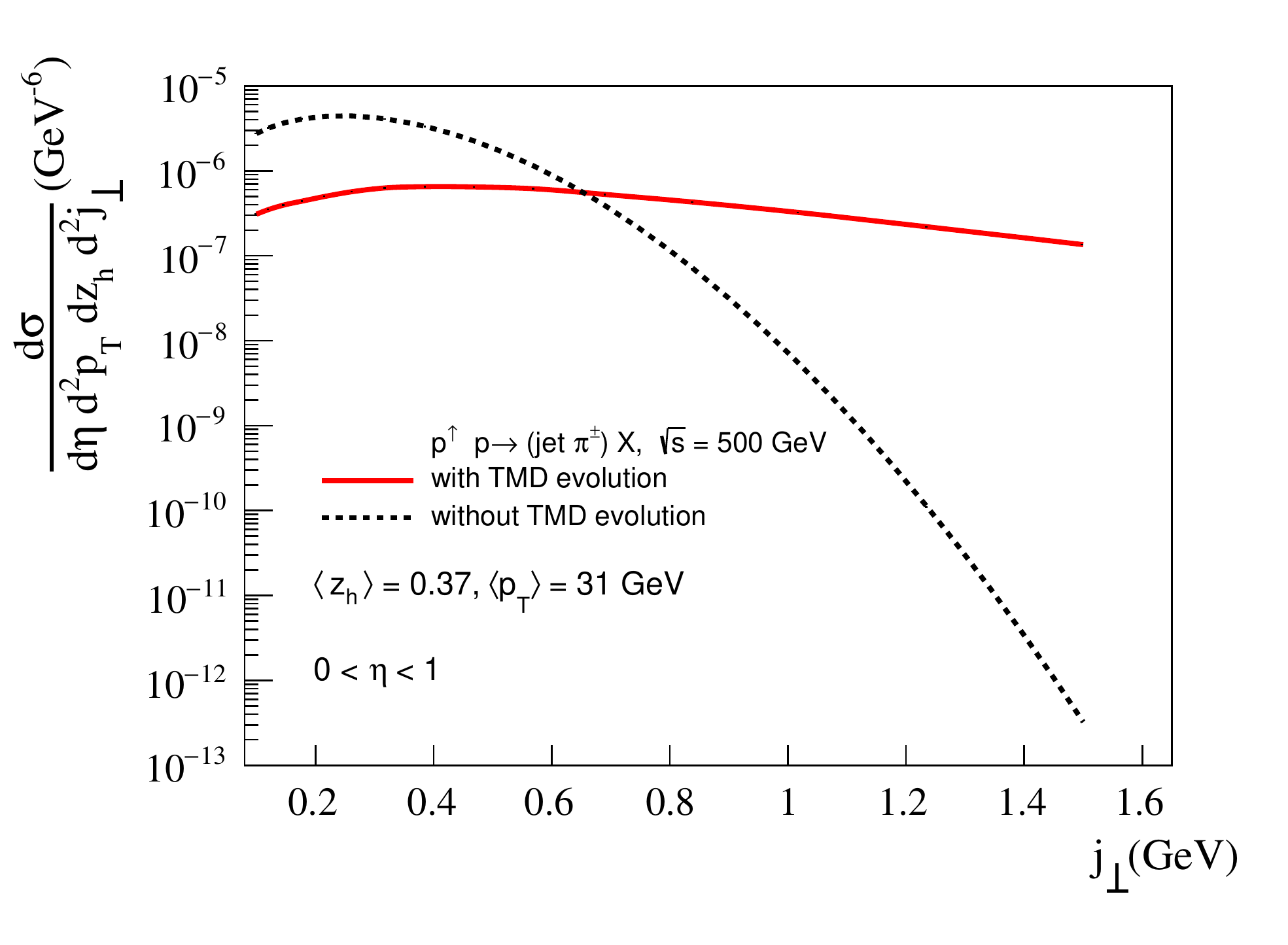}
\caption{Unpolarized $pp\to(\text{jet}\,\pi^{+})+X$ cross-section  at $\sqrt{s}=500$ GeV  as a function of the pion momentum $j_\perp$ for  $\langle z_h\rangle = 0.37$. Solid lines correspond to calculations that take into account TMD evolution using the extracted TMDs of Ref.~\cite{Kang:2015msa} and the dashed lines correspond to extraction of TMDs without TMD evolution of Ref.~\cite{Anselmino:2015sxa}. We have $\langle p_T\rangle = 31$  GeV for the average jet transverse momentum and the jet rapidity is integrated over the range $0<\eta<1$.}
\label{fig:unp_jt}
\eef

Upon completion of this work we became aware of a complementary study of the Collins azimuthal asymmetries for pion production inside jets in the framework of the Generalized Parton Model in Ref.~\cite{DAlesio:2017bvu}. The results presented in Ref.~\cite{DAlesio:2017bvu} are in agreement with our numerical results without TMD evolution and further corroborate our conclusions.

\section{Conclusion}
\label{sec:summary}
In this work, we investigated the Collins azimuthal asymmetry for hadron production inside jets in transversely polarized $p^\uparrow p$ collisions. We argued that this process is a unique opportunity to access the quark transversity distributions in the relatively large-$x$ region, and to probe the Collins fragmentation functions. In particular, the Collins fragmentation functions and the associated TMD evolution for this process are the same as those probed in the standard semi-inclusive deep inelastic scattering (SIDIS) and back-to-back di-hadron production in electron-positron annihilation. The extractions of both the quark transversity distributions and the Collins fragmentation functions from global analyses of SIDIS and electron-positron data are available in the literature with and without including TMD evolution effects. By using the extracted TMDs from these processes, we calculated the Collins azimuthal asymmetries for both positively and negatively charged pions produced inside jets $p^\uparrow p \to \left({\rm jet} \, \pi^{\pm}\right) + X$, and we compared to recent preliminary data from the STAR Collaboration at RHIC. The obtained Collins azimuthal asymmetries agree reasonably well with the experimental measurements for both CM energies $\sqrt{s} = 200$ and 500 GeV. This agreement confirms the universality of Collins fragmentation functions for the three different processes. We further explored the effects of TMD evolution, and found that the current experimental data cannot resolve the difference between our results with and without TMD evolution. We encourage the experimentalists at RHIC improve the precision of their measurements in the future, which would greatly help to assess the impact of TMD evolution effects.

\section*{Acknowledgments}
We thank J.~K.~Adkins, J.~L.~Drachenberg, R.~Fatemi and C.~A.~Gagliardi for discussions and the correspondence on the STAR experimental data. This work is supported by the Department of Energy under Contract Nos.~DE-AC05-06OR23177 (A.P.), DE-AC02-05CH11231 (F.R., F.Y.), by the National Science Foundation under Contract No.~PHY-1623454 (A.P.), by the LDRD Program of Lawrence Berkeley National Laboratory (F.R., F.Y.) and within the framework of the TMD Topical Collaboration (A.P., F.Y.).

\end{document}